# Nanosecond resolved ro-vibrational excitation of $CO_2$

Yanjun Du, Tsanko V Tsankov, Dirk Luggenhölscher and Uwe Czarnetzki

Institute for Plasma and Atomic Physics, Ruhr University Bochum, D-44780 Bochum, Germany
Email: duyanjun13@gmail.com

**Abstract**

We report first ns-resolved absorption measurements of the ro-vibrational excitation of $CO_2$. The high temporal resolution of 8 ns is made possible by a fast detector (rise-time 5 ns), sensitive in the mid-infrared region. The resolution is achieved by a slow temperature scan of a quantum cascade laser and a segmented data capturing scheme. A repetitively pulsed ns discharge in 10% $CO_2$ + 90% He at 150 mbar and a repetition rate of 2 kHz is investigated. The evolution of the population densities of the different vibration modes as well as the associated vibrational and rotational temperatures within the discharge pulse of only 150 ns length are simultaneously determined and provide valuable insight into the $CO_2$ excitation dynamics. A preferential excitation in the asymmetric vibrational mode is observed in the discharge phase shortly after the breakdown.

## Introduction

Nanosecond resolution in absorption measurements is vital for unravelling the fast temporal dynamics of the excited species in various discharges. This is especially true for repetitive nanosecond pulsed discharges, where the plasma and the entire energy deposition that drive the excitation are contained in a short pulse of a length of only a few hundreds of nanoseconds. Such discharges are investigated, among others, in relation to $CO_2$ conversion due to the large departure from equilibrium exhibited by these plasmas [1-3]. This allows a significant portion of the input energy to be transferred to the vibrational degrees of freedom of the gas molecules instead of gas heating [4-6], promoting non-equilibrium chemical reactions such as, e.g., $CO_2$ decomposition through vibrational ladder-climbing.

Previous absorption studies with ns resolution were limited to absorption by atomic species in the visible and the near-infrared region. For example, various groups [7-9] have reported He or Ar metastable populations and their temporal evolution obtained by absorption spectroscopy in nanosecond pulsed discharges. However, for understanding the physics of the discharge phase in molecular discharges, for optimizing the chemical pathways in the subsequent afterglow phase as well as for validating state-to-state kinetic models, absorption measurements on the molecular species are needed. This generally requires absorption measurements in the mid-infrared region of the spectrum and to our knowledge such measurements with ns resolution have not been reported previously. Up to now various methods for obtaining the molecular excitation, e.g., Raman scattering [10-12], IR emission spectroscopy [13, 14], Fourier transform infrared spectroscopy (FTIR) [15-18], and more recently, also tunable diode laser absorption spectroscopy (TDLAS) using narrowband lasers (~MHz) are intensively adopted to measure the $CO_2$ ro-vibrational excitation [19-23]. However, none of these methods has come close to a nanosecond resolution. For example, in a recent work of ours [23] a temporal resolution of 1.5 $\mu$s has been achieved, which was so far the common standard.

In this paper, we demonstrate the measurement strategy to monitor the broadband absorption spectrum of the $CO_2$ molecule with both high accuracy and nanosecond temporal resolution over a broad spectral range (~1 cm$^{-1}$). The method is demonstrated on a nanosecond pulsed discharge by measuring the ro-vibrational excitation of $CO_2$ within the discharge phase with a resolution of 8 ns. This is the first time when such resolution has been achieved in absorption measurements. The results offer valuable insights into the dynamics of the excited species during the discharge phase in this important case of $CO_2$ plasmas.

### Experimental setup

Measurements are performed in a nanosecond pulsed discharge in $CO_2$. The discharge configuration and the laser absorption system (figure 1) are similar to those in previous studies [2, 3, 23]. Briefly, a discharge is ignited between two parallel molybdenum electrodes 20 mm in length and spaced 1 mm apart. The discharge is well-confined and homogenous when operating with 10% $CO_2$ + 90% He at sub-atmospheric pressure of 150 mbar. It is generated by negative high-voltage (HV) pulses supplied to the electrodes by a HV switch (Behlke HTS-81). The opening and the closing of the switch is triggered by a delay generator (Stanford Research Systems DG535, marked as DDG1 in figure 1). Here, results from measurements with voltage pulses with length of 150, 200 and 250 ns are presented. The signal is monitored by a PC-based oscilloscope (PicoScope 5444B, DSO2 in figure 1). This oscilloscope, although with a moderate bandwidth of 200 MHz, provides a high vertical resolution of 14-bit and a deep buffer memory of 256 MS (for two channels). In our case the vertical resolution is important for the absorption measurement due to the large dynamic range of the laser intensity caused by the large wavelength scan. The large buffer memory is vital for achieving the high temporal resolution within a reasonable measurement time. The voltage at the powered electrode is monitored by a HV probe (Lecroy PPE6kV) and the current to ground is measured by a current probe (American Laser Systems, Model 711), both connected to another oscilloscope (Lecroy, WaveSurfer 510, bandwidth 1 GHz, 8 bit vertical resolution, marked as DSO1 in figure 1).

For the absorption a CW quantum cascade laser (Alpes) is used to scan the $CO_2$ transitions in the wavelength range between 2288.4 and 2289.6 $cm^{-1}$, enabling a simultaneous determination of the rotational and multiple vibrational temperatures for both the symmetric and the asymmetric modes [23]. After attenuation by a combination of polarizer and quarter-waveplate, the collimated light from the quantum cascade laser is guided into the discharge and finally focused by an off-axis parabolic mirror onto a fast IR detector (Vigo PVI-4TE-5, bandwidth 65 MHz). The signal from the detector is recorded by the PC-based oscilloscope (DSO2). For the characterization of the laser wavelength, a silicon etalon (LightMachinery) with a free spectral range of 0.0176 $cm^{-1}$ is introduced into the beam path by a flip mount. For a more detailed description of the other optical components the reader is referred to reference [23].

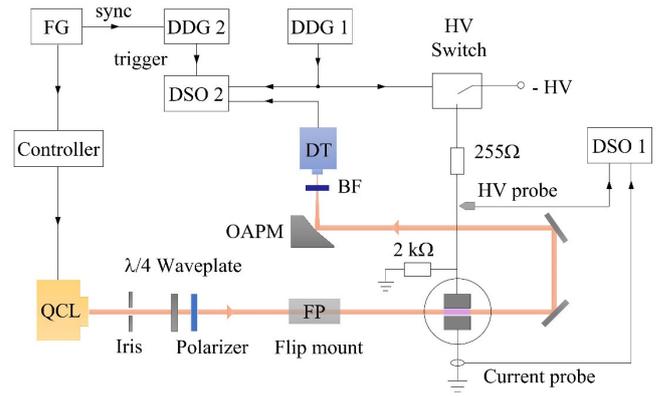

Figure 1. Schematic of the experimental setup, including the optical alignment and signal synchronization scheme for the high temporal resolution measurement. QCL: quantum cascade laser, OAPM: off-axis parabolic mirror, BF: bandpass filter, DT: detector, FP: Fabry-Perot interferometer, FG: function generator, DDG: digital delay generator, DSO: digital storage oscilloscopes, HV: high voltage.

### Measurement strategy

The ns temporal resolution is achieved by a special strategy for the laser wavelength tuning and the subsequent data acquisition (figure 2). On the laser side, a slow triangular ramp signal (frequency of $f_l$ = 10 mHz) from a function generator FG (Agilent 33250A) is sent to a commercial laser controller (ILX 3736) to scan the temperature of the quantum cascade laser within a range of $\Delta T$ = 5 °C. This temperature scanning provides a large wavelength scanning range of $\Delta v$ = 1 $cm^{-1}$ (~ - 0.2 $cm^{-1}$/°C). The temperature scanning is adopted since it provides a wider wavelength sweeping range and a smaller intensity variation compared to scanning of the laser current. Further, the continuous temperature scan has advantages over changing the temperature in steps: Better laser stability and shorter measurement times are achieved since no waiting for stabilization of the temperature setpoint is necessary. The measurements are taken during the temperature down-ramp period, which corresponds to an increasing wavenumber and generally also to an increasing laser intensity. The laser signal from the detector is recorded in steps of 8 ns. The steps are determined by the sampling rate of the PC-based oscilloscope ($S_r$ = 125 MS/s) that measures and digitizes the signal from the detector. Note that the rise-time of the detector is about 5 ns and a better oscilloscope would not improve the resolution notably. During post-processing, the recorded absorption signal is divided into time blocks of length $\Delta t$ = 20 ms (figure 2). This duration is chosen so that within each block, the temperature changes only by as much as the resolution of the controller for the temperature tuning ($\delta T = \Delta T \times f_l \times \Delta t$ = 0.001 °C). The corresponding change in the laser wavelength is $\delta v = 2 \times 10^{-4}$ $cm^{-1}$ and, therefore, the wavelength can be considered to be constant during the entire block. The approach is, therefore, termed "quasi-step-scan". Each block contains the temporal profiles of the absorption intensity from $f_d \times \Delta t = 40$ discharge pulses (for a discharge repetition frequency $f_d = 2$ kHz). Using the information from the trigger signal for the HV switch, the pulses are sorted together and averaged into a single waveform that represents the temporal evolution of the absorption for the given wavelength $v$.

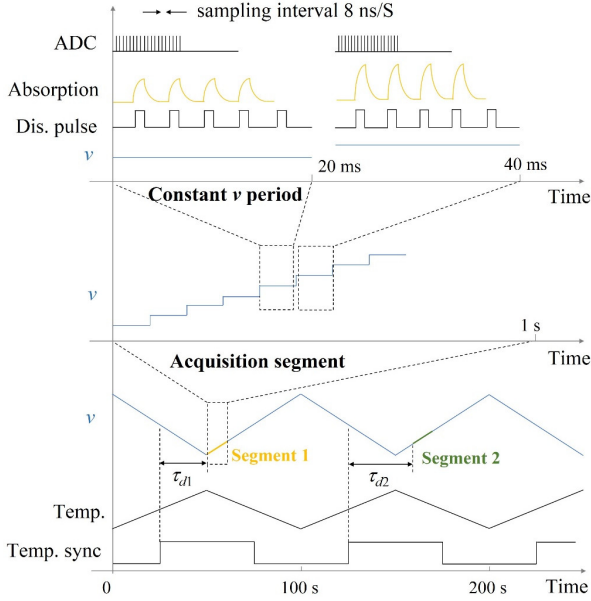

Figure 2. A timing diagram demonstrating the "quasi-step-scan" of the laser temperature and the segmented sampling scheme to achieve the ns-order temporal resolution. Bottom to top: The slow scanning ramp of the temperature and the wavelength is divided into segments, each segment is divided into 20 ms blocks, within each block the laser wavelength is (nearly) constant and the temporal evolution of 40 pulses is recorded with a resolution of 8 ns.

The obstacle here is that with such resolution the amount of data to be held in the buffer memory of the oscilloscope becomes unmanageable ($S_r / 2f_l = 6.25$ GS or about 10 GiB of data for one channel at 14 bit vertical resolution). Such an amount of data is still a challenge even for modern oscilloscopes. Therefore, in this work each temperature ramp is additionally subdivided into segments (figure 2). The length of the segments is determined by the maximal amount of data that the oscilloscope can hold in its memory (130 MS, for 14-bit and two channels).

To synchronize the laser and the segmented acquisition system, the sync signal (figure 1) of the temperature ramp from the function generator is sent to a delay generator (Stanford Research Systems DG535, DDG2) to shift the trigger of the PC-based oscilloscope (DSO2). The shift $\tau_{d,i}$ (figure 2) is incremented each period of the temperature scan by the length of one segment. With the maximum sampling rate of the oscilloscope of 125 MS/s and the maximum sampling length, 130 MS, the duration of one segment is 1.04 s and the entire temperature down-ramp of 50 s is covered by 48 segments. The dead time after the capture of each segment, needed for the completion of the data transfer from the oscilloscope (~20 s for USB2) and the data saving to the PC (~40 s in binary format), is just short enough to allow measuring the next segment already at the next period of the temperature ramp. Still for obtaining the complete temporal evolution of the spectrum with a 8 ns resolution $48/f_l = 80$ min are required.

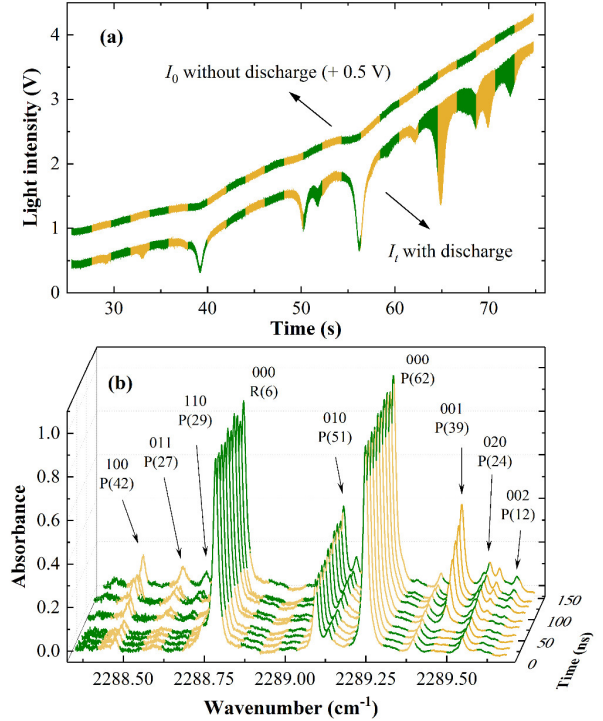

Figure 3. (a) Measured signal intensities and (b) deduced absorbance for discharge voltage of 2.25 kV and a pulse length of 150 ns. The light intensity without discharge $I_0$ is shifted vertically for clarity. The different segments are colour-coded in the spectrum. In this example the resolution was set to 16 ns in order to allow faster measurement times over a wider spectral range. The absorption peaks are labelled in the form of $v_1v_2v_3$, P- or R-branch and in brackets the rotational quantum number.

The result of this procedure is illustrated in figure 3(a) for a discharge with voltage of 2.25 kV and a pulse length of 150 ns. For this example the temporal resolution is 16 ns to allow reasonable measurement times over a wider spectral range. The laser intensity $I_0$ is obtained without a discharge and without an absorbing medium by purging the chamber with pure He. It is measured in the same manner as the intensity with plasma $I_t$, i.e. in segments and temporally resolved. The etalon signal is measured only within one single shot since no temporal resolution is needed for the wavelength calibration. After the acquisition of the different segments is complete, the signal intensity at each time instant is obtained in post-processing by first combining together the different blocks into a segment and then by "gluing" together the different segments. The temporally and spectrally resolved $CO_2$ absorbance is obtained in the usual way: $A(v, t) = \ln[I_0(v, t) / I_t(v, t)]$. The result is shown in figure 3(b). The absorbance spectra are continuous across the segments despite the corresponding $I_0$ and $I_t$ segments being measured individually in different laser scan periods. By fitting the measured time-resolved $CO_2$ absorption spectra to a simulated one [23], the evolution of the rotational and vibrational temperatures of $CO_2$ within the nanosecond discharge pulse are obtained temporally resolved.

## Results and discussions

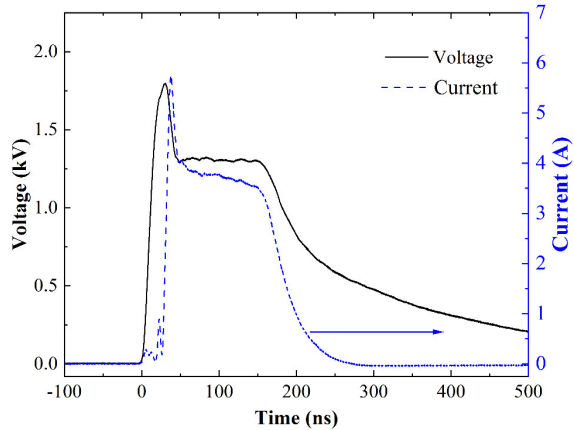

Figure 4. Voltage (solid line) and current (dashed line) waveforms in the discharge with 10% $CO_2$ + 90% He at 150 mbar, pulse length of 150 ns, repetition rate of 2 kHz and applied voltage of 2.25 kV.

The voltage and current waveforms for $CO_2$-containing discharges (figure 4) are similar to those in other molecular gases, e.g. $H_2$ [1] and $N_2$ [3]. The observed characteristic dynamics is a rapid ignition phase within the first few nanoseconds during which ionization occurs driven by a high electric field, and a subsequent quasi-DC phase, characterized by a lower quasi-stationary electric field in the discharge. For $N_2$ discharges it is shown [3] that this two-phase feature of the ns-pulsed discharges is promising for plasma-assisted chemistry. The initial strong electric field (within the first few ns) enables fast ignition and production of dense plasmas (few $10^{18}$ m$^{-3}$ for $CO_2$, value estimated from the current). The following quasi-DC phase ($t$ = 60 to 150 ns) with a moderate electric field is beneficial for the vibrational excitation while preventing significant gas heating. Based on the current and voltage waveforms, similar behaviour is expected also for plasma-assisted $CO_2$ dissociation, consistent with the results obtained here.

Figure 5 shows the time evolution of the rotational temperature, $T_{rot}$, and the vibrational temperatures, $T_{v1,v2}$, $T_{v3}$ within the discharge phase, which are obtained by fitting the time-resolved absorbance in figure 3(b) to a simulated spectrum. The model spectrum [23] uses a Treanor distribution to account for the fast vibration-vibration interactions [24, 25]. However, the vibrational temperatures reported here are essentially determined from a Boltzmann plot of the low-lying vibrational states within the scanned spectral range. The corresponding absorbing states are labelled in figure 3(b) by their vibrational quantum numbers ($v_1$ and $v_2$ for the symmetric modes and $v_3$ for the asymmetric one). Further, due to the Fermi resonances, the symmetric modes efficiently exchange energy and are modeled by a single temperature $T_{v1,v2}$.

Before the discharge pulse, $t \leq 0$ ns, the various degrees of freedom (translation, rotation, and vibration) are all in thermal equilibrium among themselves at a temperature of 320 K. This value is slightly higher than room temperature, due the energy deposited by the multiple discharge pulses. Shortly after breakdown, when an adequate number of electrons have been produced, the $CO_2$ molecules are excited and the vibrational temperatures start to increase. A pronounced excitation of the asymmetric vibrational mode is observed with a rapid increase of $T_{v3}$ from the equilibrium temperature 320 K to 738 K within the 150 ns of the discharge pulse, while only a moderate increase of just 27 K is found for the symmetric stretching and bending modes. This is accompanied by a negligible increase in the rotational temperature (considered to be equal to the gas temperature) by $\Delta T_{rot}$ = 6 K. The weak coupling of the electrons to the translational and the rotational modes of the molecules at low reduced electric field and the slow vibration-translation relaxation are the reason for this.

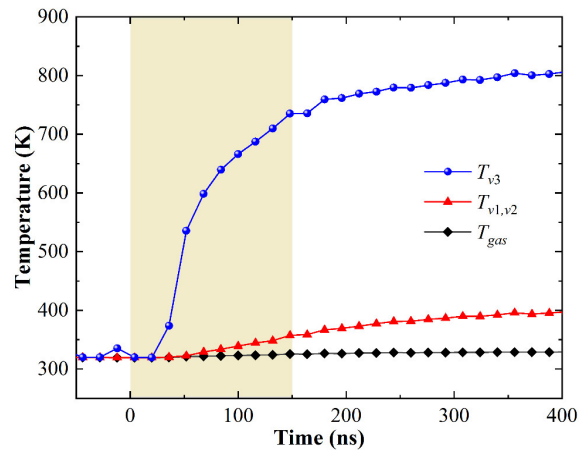

Figure 5. Temporal evolution of the best-fit gas temperatures and vibrational temperatures within the discharge phase deduced from the measured absorbance in figure 3. The extent of the discharge pulse is marked by the shaded area.

The rotational and the vibrational temperatures are only indicators for the actual ro-vibrational excitation of $CO_2$. The time-resolved populations of different vibrational levels are the more direct and valuable parameters for understanding the processes in the discharge, e.g., the so-called ladder-climbing mechanism. Figure 6 shows the temporal evolution of the densities of several vibrational states within the investigated spectral range. During the period shortly before the discharge pulse the densities of the vibrational states remain at a constant level, determined by the thermal equilibrium -- the states with higher energy have a lower density given by the Boltzmann factor. Within the discharge pulse, the density of the states associated with the asymmetric mode $v_3$, e.g. 001, 002 and 011 undergo a faster increase compared to the other vibrational states. More prominently, the first state of the asymmetric mode (001, vibrational energy of about 2350 cm$^{-1}$) is significantly faster populated than the first level of the symmetric stretching mode (100, vibrational energy of about 1330 cm$^{-1}$) demonstrating the strong departure from the equilibrium between different vibration modes of $CO_2$.

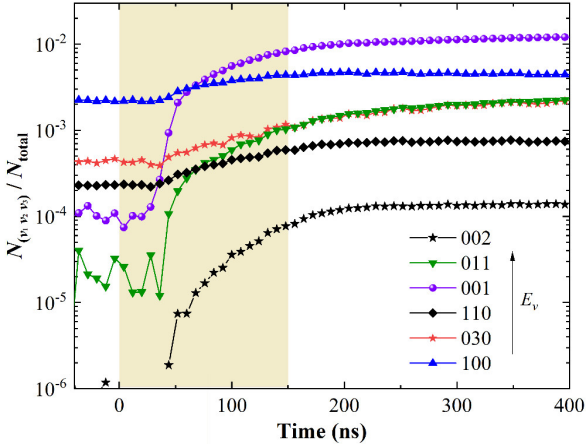

Figure 6. Time-resolved populations of the different vibrational states deduced from the scanned spectral range. The conditions are the same as in figure 4.

The ns resolution enables this first time look into the excitation dynamics during the ns discharge phase. Here we will briefly illustrate the information that this provides. The focus will be on the excitation of the asymmetric vibration of $CO_2$, relevant for efficient dissociation. The temporal evolution of the densities of the first vibrational state 001 is compared for different discharge voltages and pulse lengths (figure 7). A longer pulse length increases the period for electron excitation and results in a larger number of excited molecules, while the rate of accumulation, i.e. the slope, is not affected (figure 7a). This behaviour is captured by a simple model. The change in the density $n_{001}$ of the state 001 is due to electron impact excitation from the ground state $n_0$ (rate $k_p$) and destruction by electron collisions (net rate $k_l$):

$$\frac{dn_{001}}{dt} = k_p n_0 n_e(t) - k_l n_e(t) n_{001}(t). \quad (1)$$

Processes by heavy-particle collisions are too slow and have no effect during the discharge phase. The electron density $n_e$ is considered to be proportional to the discharge current $I(t) = eA\mu E n_e(t)$, with $e$ the elementary charge, $A$ the electrode area, $\mu$ the electron mobility and $E$ the electric field. Then the solution to equation (1) is

$$\frac{n_{001}(t)}{n_0} = \alpha \left[ 1 - \exp\left(-\beta \int_{-\infty}^{t} I(t') dt'\right) \right]. \quad (2)$$

The fit of equation (2) to the curves in figure 7a provides

$$\alpha = \frac{k_p}{k_l} = 1.98 \times 10^{-2}$$

$$\beta = \frac{k_l}{eA\mu E} = 1.2 \times 10^6 \, C^{-1}.$$

The same values are used for all conditions in figure 7. Using the mobility from BOLSIG+ [26, 27] $\mu = 0.4$ m$^2$/(V s) for the electric field measured in $N_2$ ($E$ = 100 Td) the values for the rate constants can be extracted: $k_l$ = 5.2 × 10$^{-13}$ m$^3$/s and $k_p$ = 1.0 × 10$^{-14}$ m$^3$/s. While the former is a collection of the rates for several processes out of the 001 state, the latter represents a single process and can be compared to the literature value of 0.53 × 10$^{-14}$ m$^3$/s. The value from the fit differs only by a factor of 2, despite the quite simple model approach. There can be various further causes for this difference, where one is clearly the uncertainty in the assumed electric field value. Further, there might be some partial electrode surface coverage caused by the long-term operation of the discharge during the development of this diagnostic. A lower field and/or a lower area would apparently bring the two numbers closer together.

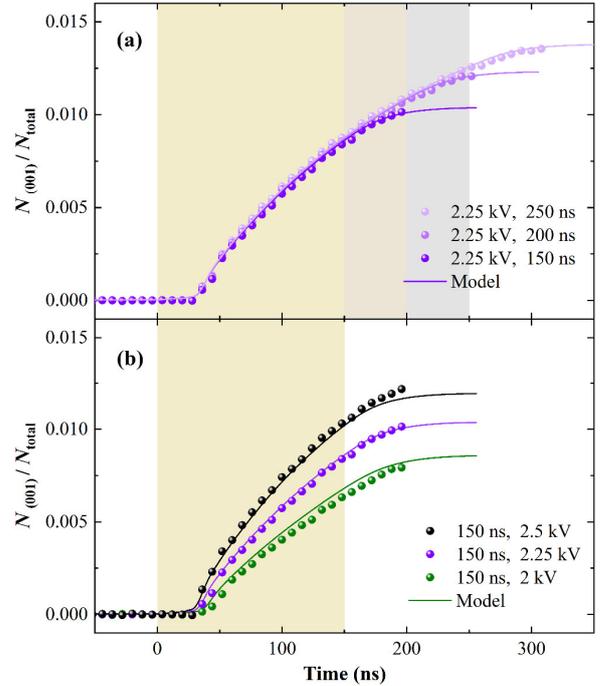

Figure 7. Densities of the first vibrational state of the asymmetric mode of $CO_2$ 001 (a) for different discharge voltages and (b) for varying pulse length. The pulse length is marked by the shaded areas. The curves are the fit of equation (2) using the same values for the parameters $\alpha$ and $\beta$.

When increasing the applied voltage a larger net production results (figure 7b). This correlates with the higher discharge currents due to a higher plasma density (a relatively weak dependence of the electric field on the applied voltage is observed in discharges in $N_2$, $N_2$/He and $H_2$ and is expected also here in $CO_2$). This is confirmed also by the very good agreement with the model from equation (2). The slight discrepancies are probably due to slight change in the electric field and, hence, in the value of *β*. A variation of 6% in *β* is sufficient to bring the agreement to the same level as for the data in figure 7a.

## Summary

In this paper, we report first measurements with ns resolution of the $CO_2$ molecular absorption in the mid-infrared region. The method for achieving this resolution together with results for the ro-vibrational excitation of $CO_2$ in a nanosecond pulsed discharge are presented. The time-resolved evolution of the rotational and the vibrational temperatures of $CO_2$ within the nanosecond discharge pulse show a preferential excitation of the asymmetric stretch mode and the large departure from equilibrium caused by the plasma. The evolution of the density of the first state of the asymmetric vibrational mode is explained by a simple model that allows the determination of the excitation rates.

## Acknowledgements

This work is supported by the DFG in the frame of SFB1316 Project "Transient atmospheric plasmas: from plasmas to liquids to solids". Yanjun Du also acknowledges the financial support from the Alexander von Humboldt Foundation.